\documentclass{ws-procs961x669}

\usepackage{afterpage}
\usepackage{hyperref}
\usepackage{longtable}
\usepackage{threeparttablex}

\newcommand{\mej}{$M_{\textrm{ej}}$}
\newcommand{\spin}{$P_\textrm{i}$}

\begin{document}

\title{\title{Light Curve Properties of Gamma-Ray Burst Associated Supernovae}}

\author{Amit Kumar}

\address{Department of Physics, Royal Holloway, University of London, TW20 0EX, UK\\
Department of Physics, University of Warwick, Coventry, CV4 7AL, UK\\
$^*$E-mail: amitkundu515@gmail.com, amit.kumar@rhul.ac.uk\\}

\author{Kaushal Sharma}

\address{Forensic Science Laboratory Uttar Pradesh, Moradabad 244 001, India\\
Inter University Centre for Astronomy and Astrophysics (IUCAA), Pune 411 007, India\\
E-mail: kaushal.28838@up.gov.in}

\begin{abstract}
A rapidly spinning, millisecond magnetar is widely considered one of the most plausible power sources for gamma-ray burst-associated supernovae (GRB-SNe). Recent studies have demonstrated that the magnetar model can effectively explain the bolometric light curves of most GRB-SNe. In this work, we investigate the bolometric light curves of 13 GRB-SNe, focusing on key observational parameters such as peak luminosity, rise time, and decay time, estimated using Gaussian Process (GP) regression for light curve fitting. We also apply Principal Component Analysis to all the light curve parameters to reduce the dimensionality of the dataset and visualize the distribution of SNe in lower-dimensional space. Our findings indicate that while most GRB-SNe share common physical characteristics, a few outliers, notably SNe 2010ma and 2011kl, exhibit distinct features. These events suggest potential differences in progenitor properties or explosion mechanisms, offering deeper insight into the diversity of GRB-SNe and their central engines.
\end{abstract}


\bodymatter

\section{Introduction}\label{sec:intro}

The connection between long gamma-ray bursts (LGRBs) and broad-lined H/He-deficient supernovae (Type Ic-BL SNe) was first established with the discovery of the nearby GRB 980425, which was accompanied by SN 1998bw \citep{Galama1998b, Iwamoto1998, Wang1998, Patat2001}. Over the past two and a half decades, this field has evolved significantly, with the identification of more than fifty confirmed GRB-SN events \cite{Woosley2006, Cano2017, Dainotti2022, Aimuratov2023, Li2023, Kumar2024}. Among these, SN 2011kl stands out as the only superluminous SN (SLSN), associated with the ultra-long (ul) GRB 111209A \cite{Greiner2015, Gompertz2017, Kann2019}; however, only a handful of GRB-SNe have been well-observed across multiple bands. See the recent work by Ref.~\citenum{Finneran2024} for detailed data and a comprehensive compilation of parameters for GRBs and their associated SNe.

This scarcity of well-observed GRB-SNe stems from several factors beyond the relatively low occurrence of GRB-SNe. These include the lower observed flux from SNe connected to high-redshift GRBs and dust extinction within the interstellar medium and host galaxies, particularly in the blue wavelengths, which hinders SN detection \cite{Holland2010}. Furthermore, not all GRBs produce bright SNe, possibly due to insufficient $^{56}$Ni production, significant fallback onto the newly formed black hole, or lower-energy explosions \cite{Tominaga2007}, also see Refs.~\citenum{Kumar2022d, Shrestha2023}. Thus, an in-depth investigation of the limited sample of GRB-SNe is crucial for understanding their unique properties, and exploring their power sources remains one of the important keys to unlocking their nature.

A widely accepted model suggests that a newborn millisecond magnetar, formed during the collapse of a massive star, could serve as the power source for both GRBs and their associated SNe \cite{Usov1992, Wheeler2000, Woosley2010, Mazzali2014, Metzger2015, Kashiyama2016, Margalit2018, Inserra2019, Shankar2021, Ho2023, Kumar2024, Omand2024}. Several studies have proposed the magnetar as the central engine in individual cases or small samples of GRB-SNe \cite{Soderberg2006a, Margutti2013, Greiner2015, Barnes2018, Kumar2022a}. Additionally, millisecond magnetars have been proposed as power sources to explain the properties of various other types of SNe, including SLSNe \cite{Quimby2011, Inserra2013, Nicholl2017, Yu2017, Dessart2019, Kumar2020, Kumar2021ank}, classical Ic-BL SNe (e.g., SNe~1997ef and 2007ru \cite{Wang2016}), Type Ic (e.g., SN 2019cad \cite{Gutierrez2021}), Type Ib (e.g., SN~2005bf \cite{Maeda2007a} and SN~2012u \cite{Pandey2021, Omand2023}), and fast blue optical transients (FBOTs \cite{Hotokezaka2017, Prentice2018, Fang2019, Liu2022}). 

More recently, through light curve modelling of all GRB-SNe with comprehensive multi-band data, Ref.~\citenum{Kumar2024} (hereafter {\tt K24}) demonstrated that the magnetar model can successfully explain the light curves of nearly all observed GRB-SNe. This study also explored magnetar parameters across a range of cosmic transients—including SLSNe, FBOTs, and both long and short GRBs$-$showing that variations in parameters like magnetic field strength and initial spin period can lead to the formation of different types of transients. In this study, we investigate the diversity of bolometric light curve parameters of 13 GRB-SNe in multi-dimensional space using parameters reported in {\tt K24} along with some newly derived parameters for 13 GRB-SNe.

\begin{figure}[h]
\begin{center}
\includegraphics[width=1\textwidth]{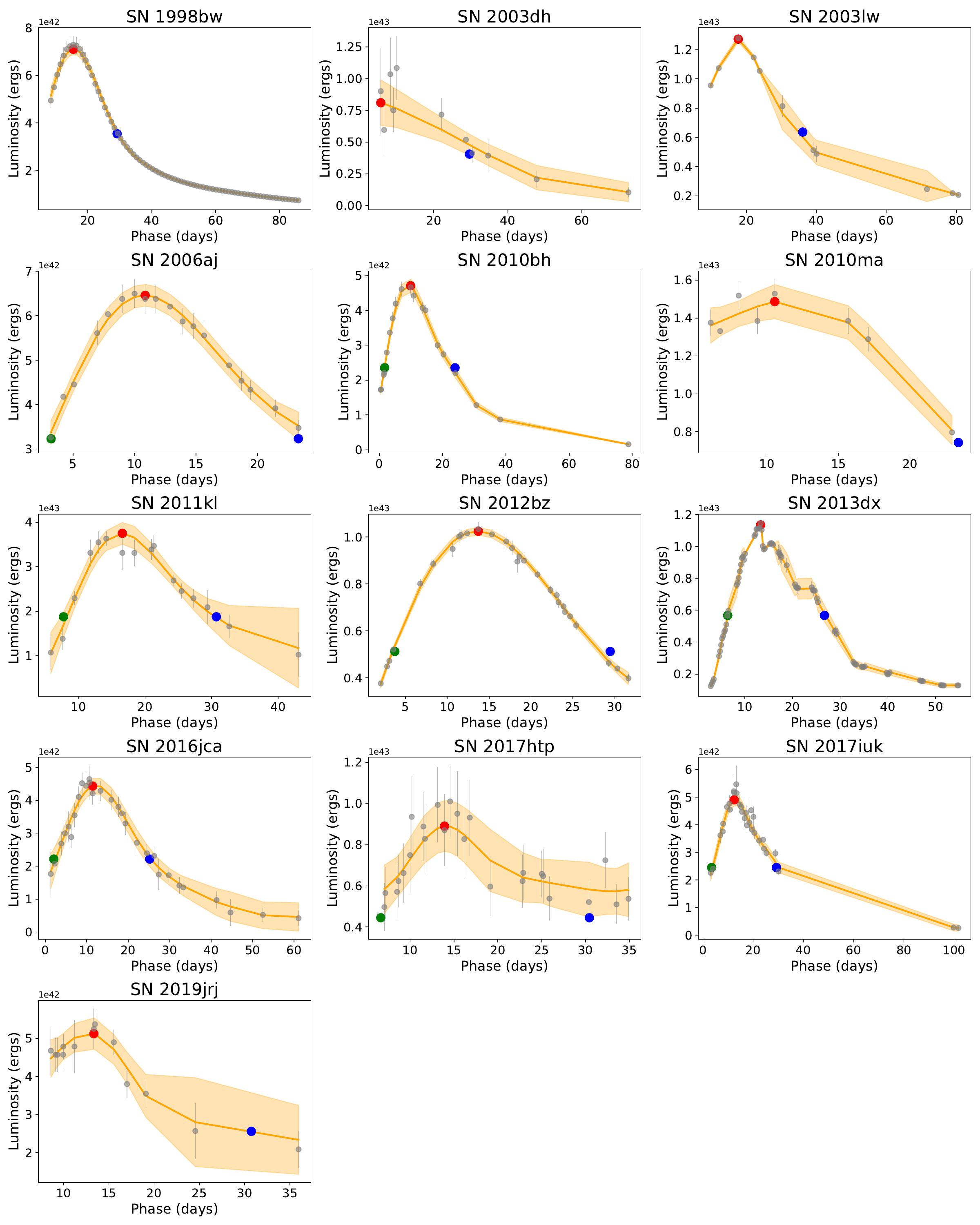}
\end{center}
\caption{Bolometric light curves of all 13 GRB-SNe along with GP fits are shown. The phases corresponding to L$_p$, L$p_{/2}$ in pre-peak data, and L$p_{/2}$ in post-peak data are shown by the red, green and blue dots, respectively.}
\label{fig:lcs}
\end{figure}

\section{Bolometric Light Curves}\label{sec:lcs}

The GRB-SNe used in the present work, along with the names of their associated GRBs (in parentheses), are as follows: SN~1998bw (GRB~980425), SN~2003dh (GRB~030329), SN~2003lw (GRB~031203), SN~2006aj (GRB~060218), SN~2010bh (XRF~100316D), SN~2010ma (GRB~101219B), SN~2011kl (GRB~111209A), SN~2012bz (GRB~120422A), SN~2013dx (GRB~130702A), SN~2016jca (GRB~161219B), SN~2017htp (GRB~171010A), SN~2017iuk (GRB~171205A), and SN~2019jrj (GRB~190114C). The bolometric light curves for all 13 GRB-SNe, listed in Table~\ref{tab:param} and analyzed in this study were directly taken from {\tt K24}, which compiled these data from Refs.~\citenum{Cano2017, Cano2017a, Postigo2017, Wang2018, Ashall2019, Izzo2019, Melandri2019, Suzuki2019, Kumar2022a, Melandri2022} and references therein. For details regarding the sample selection criteria, the method used to extract SN light curves from GRB+SN+host contributions, and the construction of bolometric light curves, refer to {\tt K24}. Additionally, light curve modelling of GRB-SNe using the {\tt MINIM} code \cite{Chatzopoulos2013}, assuming a millisecond magnetar as the powering source, was performed in {\tt K24}, constraining various parameters such as initial rotational energy of the magnetar ($E_\textrm{p}$), diffusion timescale ($t_\textrm{d}$), spin-down timescale ($t_\textrm{p}$), progenitor star radius ($R_\textrm{p}$), ejecta expansion velocity ($V_\textrm{exp}$), ejecta mass ($M_\textrm{ej}$), initial spin period ($P_\textrm{i}$), and magnetic field strength ($B$).

As an extension of the analysis conducted in {\tt K24}, in this study, we estimate the peak luminosity ($L_p$), peak time ($t_r$, time from explosion to peak luminosity, where the explosion time corresponds to the GRB detection time by satellite), rise time (time from half of the peak luminosity to peak in pre-peak data, $t_{r_{L/2}}$), and decay time (time from peak to half of the peak luminosity in post-peak data, $t_{d_{L/2}}$). To estimate these parameters, we used $t_{r_{L/2}}$ and $t_{d_{L/2}}$, which effectively capture the light curve evolution around the peak.

To estimate $L_p$, $t_r$, $t_{r_{L/2}}$ and $t_{d_{L/2}}$, the light curve fitting was performed using Gaussian Processes (GP) regression \cite{Rasmussen2005, Bishop2006} with a Radial Basis Function (RBF) kernel for interpolation, providing a more accurate fit compared to spline fitting. This approach replicates the bolometric light curves while naturally providing uncertainty estimates as a function of phase \cite{Inserra2018a}. We performed these analyses using Python packages {\tt sklearn} and {\tt scipy}.

\begin{sidewaystable}[!p]
\tbl{Light curve fitting parameters considering magnetar as a possible powering source adopted from {\tt K24} along with $L_p$, $t_r$, $t_{r_{L/2}}$ and $t_{d_{L/2}}$.}
{\begin{tabular}{@{\hskip 0cm}c@{\hskip 0cm}c@{\hskip 0.1cm}c@{\hskip 0.35cm}c@{\hskip 0.35cm}c@{\hskip 0.35cm}c@{\hskip 0.35cm}c@{\hskip 0.35cm}c@{\hskip 0.35cm}c@{\hskip 0.35cm}c@{\hskip 0.35cm}c@{\hskip 0.35cm}c@{\hskip 0.35cm}c@{\hskip 0.35cm}c@{\hskip 0.35cm}}
\hline
SN & Associated & $E_\textrm{p}^a$ & $t_\textrm{d}^b$ & $t_\textrm{p}^c$ & $R_\textrm{p}^d$ & $V_\textrm{exp}^e$ & \mej{}$^f$ & \spin{}$^g$ & $B^h$ & L$_p^i$ & t$_r^j$ & $t_{r_{L/2}}^k$ & $t_{d_{L/2}}^l$ \\
 & GRB & ($10^{49}$~erg) & (d) & (d) & (10$^{13}$~cm) & (10$^{3}$~km s$^{-1}$) & ($M_{\odot}$) & (ms) & ($10^{14}$~G) & ($10^{43}$~erg/s)  & days & days & days \\ 
\hline \hline
1998bw & 980425 &  2.67  $\pm$ 0.01  & 12.35 $\pm$  0.07   &  9.49  $\pm$  0.09 &  19.26  $\pm$  0.57 &  30.81  $\pm$  1.43 &  3.62  $\pm$  0.21  &  27.38  $\pm$  0.04 &   19.38  $\pm$  0.09&   0.72  $\pm$  0.008 & 15.68  $\pm$  0.96 &  nan  $\pm$   nan  &  13.69  $\pm$  0.99 \\
2003dh & 030329 & 4.76  $\pm$ 0.60  & 16.32 $\pm$  2.01   &  9.44  $\pm$  1.49 &  11.41  $\pm$  2.35 &  22.21  $\pm$  8.57 &  4.55  $\pm$  2.88  &  20.50  $\pm$  1.30 &   14.56  $\pm$  1.15&   0.88  $\pm$  0.06 & 7.36   $\pm$  1.53 &  nan  $\pm$   nan  &  24.04  $\pm$  2.92 \\
2003lw & 031203 & 3.62  $\pm$ 0.19  & 19.77 $\pm$  0.79   &  4.90  $\pm$  0.21 &  84.67  $\pm$  6.75 &  20.59  $\pm$  1.76 &  6.20  $\pm$  1.02  &  23.51  $\pm$  0.62 &   23.17  $\pm$  0.49&   1.27  $\pm$  0.01 & 17.64  $\pm$  0.00 &  nan  $\pm$   nan  &  18.72  $\pm$  4.33 \\
2006aj & 060218 & 10.54 $\pm$ 0.72  & 18.12 $\pm$  1.21   &  0.52  $\pm$  0.05 &  0.001  $\pm$  0.36 &  20.47  $\pm$  6.39 &  5.18  $\pm$  2.31  &  13.78  $\pm$  0.47 &   41.58  $\pm$  2.18&   0.66  $\pm$  0.009 & 10.74  $\pm$  0.81 &  7.51 $\pm$   0.77 &  12.49  $\pm$  0.92 \\
2010bh & 100316D & 2.89  $\pm$ 0.03  & 15.79 $\pm$  0.80   &  3.69  $\pm$  0.10 &  0.04   $\pm$  0.13 &  29.54  $\pm$  2.97 &  5.67  $\pm$  1.15  &  26.32  $\pm$  0.16 &   29.87  $\pm$  0.41&   0.47  $\pm$  0.008 & 9.83   $\pm$  0.98 &  8.23 $\pm$   0.98 &  13.98  $\pm$  1.41 \\
2010ma & 101219B & 17.75 $\pm$ 1.96  & 18.41 $\pm$  0.51   &  0.48  $\pm$  0.19 &  7.16   $\pm$  1.02 &  15.32  $\pm$  3.67 &  4.00  $\pm$  1.18  &  10.61  $\pm$  0.59 &   33.23  $\pm$  6.65&   1.50  $\pm$  0.04 & 10.07  $\pm$  1.09 &  nan  $\pm$   nan  &  12.87  $\pm$  1.09 \\
2011kl & 111209A & 11.87 $\pm$ 0.71  & 12.68 $\pm$  0.32   &  12.70 $\pm$  1.22 &  10.26  $\pm$  3.56 &  24.46  $\pm$  2.46 &  3.03  $\pm$  0.46  &  12.98  $\pm$  0.39 &   7.95   $\pm$  0.38&   3.79  $\pm$  0.11 & 16.86  $\pm$  1.21 &  8.85 $\pm$   1.39 &  14.04  $\pm$  3.37 \\
2012bz & 120422A & 7.30  $\pm$ 0.22  & 19.15 $\pm$  0.13   &  3.59  $\pm$  0.14 &  8.44   $\pm$  1.18 &  28.87  $\pm$  4.72 &  8.15  $\pm$  1.44  &  16.56  $\pm$  0.25 &   19.06  $\pm$  0.38&   1.03  $\pm$  0.007 & 13.56  $\pm$  0.82 &  10.0 $\pm$1  0.82 &  15.73  $\pm$  0.82 \\
2013dx & 130702A & 4.36  $\pm$ 0.05  & 13.65 $\pm$  0.04   &  5.38  $\pm$  0.04 &  10.62  $\pm$  1.67 &  23.52  $\pm$  4.88 &  3.37  $\pm$  0.72  &  21.42  $\pm$  0.11 &   20.13  $\pm$  0.08&   1.14  $\pm$  0.003 & 13.29  $\pm$  0.08 &  6.91 $\pm$   0.13 &  13.44  $\pm$  1.86 \\
2016jca & 161219B & 2.34  $\pm$ 0.08  & 14.96 $\pm$  0.57   &  7.37  $\pm$  0.82 &  1.86   $\pm$  0.50 &  33.89  $\pm$  3.91 &  5.84  $\pm$  1.12  &  29.25  $\pm$  0.51 &   23.50  $\pm$  1.31&   0.45  $\pm$  0.008 & 11.68  $\pm$  1.18 &  9.40 $\pm$   1.36 &  13.72  $\pm$  1.59 \\
2017htp & 171010A & 5.43  $\pm$ 0.01  & 17.15 $\pm$  0.33   &  13.94 $\pm$  1.50 &  0.07   $\pm$  0.03 &  14.37  $\pm$  0.66 &  3.25  $\pm$  0.27  &  19.18  $\pm$  0.20 &   11.21  $\pm$  0.61&   0.96  $\pm$  0.04 & 14.40  $\pm$  1.29 &  7.26 $\pm$   1.20 &  16.40  $\pm$  4.08 \\
2017iuk & 171205A & 2.64  $\pm$ 0.01  & 17.01 $\pm$  1.26   &  10.91 $\pm$  0.41 &  0.05   $\pm$  0.02 &  28.72  $\pm$  4.08 &  6.40  $\pm$  1.86  &  27.50  $\pm$  0.05 &   18.16  $\pm$  0.34&   0.50  $\pm$  0.008 & 12.87  $\pm$  0.72 &  8.90 $\pm$   0.76 &  16.80  $\pm$  0.95 \\
2019jrj & 190114C & 6.14  $\pm$ 0.28  & 19.51 $\pm$  1.88   &  1.09  $\pm$  0.02 &  0.51   $\pm$  0.07 &  23.93  $\pm$  3.71 &  7.01  $\pm$  2.44  &  18.04  $\pm$  0.41 &   37.75  $\pm$  0.29&   0.53  $\pm$  0.02 & 12.92  $\pm$  1.08 &  nan  $\pm$   nan  &  17.02  $\pm$  6.32 \\
\Hline
Median & values: & 4.76 $\pm$ 0.07 & 17.01 $\pm$ 0.20 & 5.38 $\pm$ 0.07 & 7.16 $\pm$ 0.20 & 23.93 $\pm$ 1.29 & 5.18 $\pm$ 0.40 & 20.50 $\pm$ 0.14 & 20.13 $\pm$ 0.12 & 0.88  $\pm$  0.01 & 12.92  $\pm$  0.34 &  8.54 $\pm$ 0.44  &  13.98  $\pm$  0.65 \\
\hline
\end{tabular}}\label{tab:param}
    \begin{tablenotes}[para,flushleft]
        $^a$ $E_\textrm{p}$: initial rotational energy of the magnetar ($10^{49}$~erg).
        $^b$ $t_\textrm{d}$: effective diffusion timescale (days).
        $^c$ $t_\textrm{p}$: spin-down timescale (days).    
        $^d$ $R_\textrm{p}$: progenitor star's radius (10$^{13}$~cm).
        $^e$ $V_\textrm{exp}$: expansion velocity (10$^{3}$~km s$^{-1}$).
        $^f$ \mej: ejected mass ($M_\odot$).
        $^g$ \spin: initial spin period (ms).
        $^h$ $B$: magnetic field ($10^{14}$~G).
        $^i$ $L_p$: peak luminosity (10$^{43}$~erg s$^{-1}$).
        $^j$ $t_r$: time between explosion and peak-luminosity (days).
        $^k$ $t_{r_{L/2}}$: time between peak-luminosity/2 and peak-luminosity in pre-peak phase (days).
        $^l$ $t_{d_{L/2}}$: time between peak-luminosity and peak-luminosity/2 in post-peak phase (days).
    \end{tablenotes}
\end{sidewaystable}

The bolometric light curves of all 13 GRB-SNe in this study, along with GP interpolation for each, are shown in Figure~\ref{fig:lcs}. The phases where $L_p$, $L_{p_{/2}}$ in the pre-peak, and $L_{p_{/2}}$ in the post-peak data are located, are marked by red, green, and blue dots on the GP fits, respectively. Due to insufficient pre-peak data, we were unable to calculate $t_{r_{L/2}}$ for SNe 1998bw, 2003dh, 2003lw, 2010ma, and 2019jrj. 

The estimated parameters $L_p$, $t_r$, $t_{r_{L/2}}$, and $t_{d_{L/2}}$ for all GRB-SNe, along with their medians, are provided in Table~\ref{tab:param}, alongside other light curve fitting parameters adopted from table 1 of {\tt K24}. The median values of the parameters determined for the GRB-SNe in our sample are as follows: $E_\textrm{p} \approx 4.8 \times 10^{49}$ erg, $t_\textrm{d} \approx 17$ days, $t_\textrm{p} \approx 5.4$ days, $R_\textrm{p} \approx 7.2 \times 10^{13}$ cm, $V_\textrm{exp} \approx 24,000$ km s$^{-1}$, $M_\textrm{ej} \approx 5.2 M_\odot$, $P_\textrm{i} \approx 20.5$ ms, $B \approx 20.1 \times 10^{14}$ G, $L_p \approx 0.88 \times 10^{43}$ erg s$^{-1}$, $t_r \approx 13$ days, $t_{r_{L/2}} \approx 8.5$ days, and $t_{d_{L/2}} \approx 14$ days.

\section{Dimensionality Reduction}
To explore the estimated parameters among the GRB-SNe in this study, we apply Principal Component Analysis (PCA) to the parameters listed in Table~\ref{tab:param}, aiming to reduce the dimensionality of the dataset and visualize the distribution of SNe in lower-dimensional space.

PCA transforms a multi-dimensional parameter space into a set of orthogonal components, called Principal Components (PCs). These components are linear combinations of the original parameters and are constructed to maximize the variance captured in the data while minimizing redundancy. PCA helps simplify complex datasets by identifying the most important directions of variation, allowing us to reduce the number of dimensions while preserving as much information as possible for analysis and visualization.

\begin{figure}[h]
\begin{center}
\includegraphics[width=1.0\textwidth]{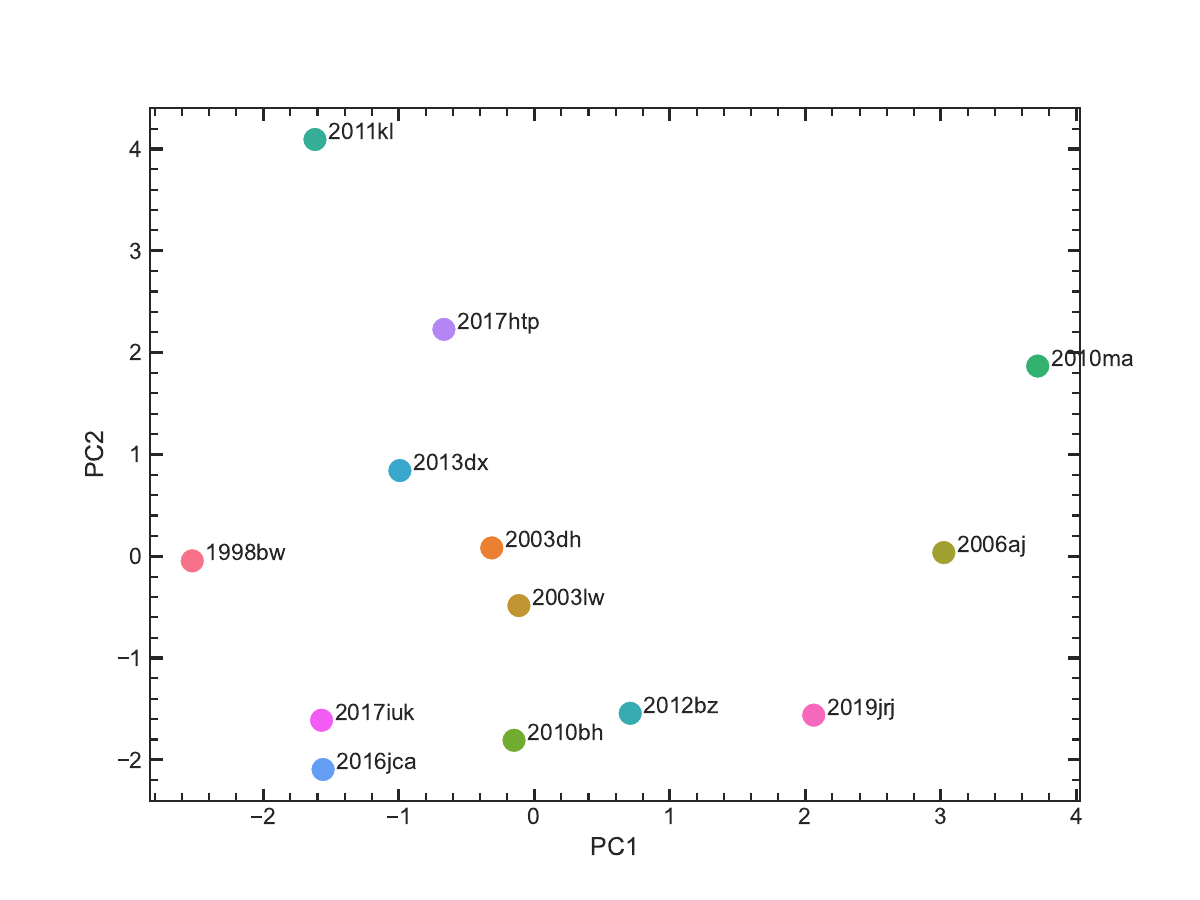}
\end{center}
\caption{Distribution of SNe in the space of the first two principal components (PC1 and PC2), accounting for approximately 50\% of the variance in the dataset. Each point represents a SN, colour-coded by name, and labelled accordingly.}
\label{fig:pca_distribution}
\end{figure}

PCA analysis reveals that the first two principal components (PC1 and PC2) capture approximately 50\% of the variance in the dataset, with PC1 accounting for 25.5\% and PC2 for 24.4\%. The first five components collectively explain about 90\% of the variance in the data, which is sufficient for visualizing and interpreting key trends. 
Figure~\ref{fig:pca_distribution} shows the distribution of the 13 SNe projected onto the PC1-PC2 plane. Most of the SNe form a relatively compact group near the origin, suggesting that their physical parameters share commonalities. 

SNe 2019jrj and 2006aj appear somewhat isolated, particularly along PC1, while SN 2017htp stands out along PC2, indicating that these objects may possess distinct parameter characteristics, e.g., SN 2017htp exhibits the highest spin-down timescale and SN 2006aj presents the lowest progenitor radius and highest magnetic field. Notably, SNe 2011kl and 2010ma, deviate significantly from the main cluster, suggesting that these SNe exhibit distinct observational and physical characteristics. As highlighted in {\tt K24}, SN 2011kl (one and only known ulGRB-associated SLSN) and SN 2010ma show the highest peak luminosity and magnetar initial rotational energy among the sample of GRB-SNe (see Table~\ref{tab:param}). Furthermore, SN 2011kl is characterized by the lowest magnetic field strength, while SN 2010ma displays the shortest spin-down timescale compared to all other events in the dataset. In addition to the diversity in the powering source parameters, this diversity in properties can likely be attributed to differences in their progenitor properties and host environment. These deviations underscore the complexity and diversity within the population of GRB-SNe, providing key insights into the diverse nature of their progenitors and explosion dynamics.

\section{Discussion and Conclusion}

In this study, we analyzed the bolometric light curves of 13 GRB-SNe and examined key physical and observational parameters. For each SN, we derived important light curve parameters such as peak luminosity ($L_p$), rise time ($t_r$), and both the pre-peak ($t_{r_{L/2}}$) and post-peak ($t_{d_{L/2}}$) half-luminosity times using GP regression with an RBF kernel. This approach provided precise interpolations and uncertainty estimates, enabling a more robust analysis of the light curves and their associated parameters.

We applied PCA to the light curve parameters estimated in this study, alongside those adopted from {\tt K24}, to explore correlations and reduce dimensionality. The PCA revealed that most SNe are clustered near the origin of the PCA plot, indicating shared physical characteristics. However, a few outliers, such as SNe 2010ma and 2011kl, diverged from the main cluster, suggesting distinct observational and physical properties. These deviations could also be attributed to the differences in progenitor properties and environments of these SNe.

These findings underscore the need for more advanced modelling and theoretical studies to fully capture the diverse nature of GRB-SNe and their central engine-based powering sources. The results also emphasize the value of statistical techniques like PCA in unravelling the complex multi-dimensional parameter space of SNe. Expanding the sample size of GRB-SNe in future studies will allow for more rigorous statistical analysis, further enhancing our understanding of these extraordinary cosmic events.

\section*{Acknowledgement}
A.K. is supported by the UK Science and Technology Facilities Council (STFC) Consolidated grant ST/V000853/1. This research has made use of data from the NASA/IPAC Extragalactic Database (NED), which is operated by the Jet Propulsion Laboratory at the California Institute of Technology, under contract with NASA. We also acknowledge the invaluable support of NASA's Astrophysics Data System Bibliographic Services.

\bibliographystyle{ws-procs961x669}
\bibliography{ws-pro-sample}

\end{document}